\documentclass[epj]{svjour}
\usepackage{graphicx}
\usepackage{epsfig}
\begin{document}
\title{Recursion method and one-hole spectral function of the 
Majumdar-Ghosh model}
\author{R. O. Kuzian\inst{1} \and R. Hayn \inst{2}$^,$\inst{3} \and 
J. Richter\inst{4}}
\institute{
Institute for Materials Science, Ukrainian Academy of
Sciences,
Krzhizhanovskogo 3, 03180 Kiev, Ukraine \and
Leibniz-Institut f\"ur Festk\"orper- und 
Werkstoffforschung Dresden,
P.O.Box 270016, D-01171 Dresden, Germany \and
Laboratoire Mat\'eriaux et Micro\'electronique de Provence, 
49, rue Joliot-Curie, IRPHE, F-13384 Marseille Cedex 13, France
\and
Institut f\"ur Theoretische Physik, 
Otto-von-Guericke Universit\"at Magdeburg, 
P.O.Box 4120, D-39016 Magdeburg, Germany }
\date{\today }

\abstract{
We consider the application of the recursion method to the calculation of
one-particle Green's functions for strongly correlated systems and propose a
new way how to extract the information about the infinite system from the
exact diagonalisation of small clusters. Comparing the results for several
cluster sizes allows us to establish those Lanczos coefficients that are not
affected by the finite size effects and provide the information about the
Green's function of the macroscopic system. The analysis of this
'bulk-related' subset of coefficients supplemented by alternative analytic
approaches allows to infer their asymptotic behaviour and to propose an
approximate analytical form for the 'terminator' of the Green's function
continued fraction expansion for the infinite system. As a result, the
Green's function acquires the branch cut singularity corresponding to the
incoherent part of the spectrum. The method is applied to the spectral
function of one-hole in the Majumdar-Ghosh model (the one-dimensional $
t-J-J^{\prime}$ model at $J^{\prime }/J=1/2$). For this model, the branch
cut starts at finite energy $\omega_0$, but there is no upper bound of the
spectrum, corresponding to a linear increase of the recursion coefficients.
Further characteristics of the spectral function are band gaps in the middle
of the band and bound states below $\omega_0$ or within the gaps. The band
gaps arise due to the period doubling of the unit cell and show up as
characteristic oscillations of the recursion coefficients on top of the
linear increase.
} 

\maketitle

\section{Introduction}

The one-hole spectral function (or the imaginary part of the one-particle
Green's function) is a central quantity in the theory of strongly correlated
systems since it contains the information about the charge dynamics in a
correlated background \cite{Fulde}. It was often studied in the theory of
high-temperature superconductors (HTSC) using the several variants of the $t$
-$J$ or the Hubbard model \cite{Dagotto}. Other examples are the Kondo peak
seen in the spectral function of heavy fermion compounds \cite{Hewson}, or
the features of spin-charge separation for one-dimensional problems 
\cite{Voit}. The most important experimental method to measure the one-hole
spectral function is angle-resolved photoemission.

A powerful method to calculate the one-hole spectral function for strongly
correlated systems consists in the exact diagonalisation of small clusters 
\cite{Dagotto} using the Lanczos method \cite{Lanczos}. That can be
understood like a tridiagonalisation of the Hamilton matrix. The method and
its application in electronic structure theory has already a long history.
It is also known as recursion method according to pioneering studies of
Haydock and coworkers in the seventies \cite{Haydock72,Haydock80}. That
time, however, it was mostly applied to different problems of noninteracting
electrons. The numerous spectral function calculations using the different
versions of the $t$-$J$ or the Hubbard model \cite{Dagotto} have usually
very large finite size effects. Normally, they provide the spectral function
only as a sequence of well separated $\delta $-peaks. On the other hand, it
is already known for long time that the recursion method has the potential
to provide the information about the bulk spectral function by a proper
termination of the continued fraction representation of the Green's function 
\cite{HayNex85,Ducastelle}. That idea was used in the field of many-body
dynamics for pure spin systems by Viswanath and M\"uller \cite{VisMul}.

Below we would like to present an improvement of the recursion method for
the charge dynamics in strongly correlated systems. By finding the correct
asymptotic behaviour of the recursion coefficients $a_n$ and $b_n$ for $n\to
\infty $ which is not influenced by boundary effects we are able to extract
the bulk spectral function. The method will be demonstrated for the example
of one-hole in the Majumdar-Ghosh (MG) model, i.e.\ in the 1D spin chain
with exchange terms between first $J$ and second neighbours $J^{\prime }$ at
the special ratio $J^{\prime }/J=1/2$. It has the advantage that one knows
explicitly the ground-state wave function of the spin system \cite{Majumdar}. 
The MG model is considered here as a generic model for a 1D spin chain
with a spin gap. That means that the spin correlation functions decay
exponentially. Therefore, the MG model is a good candidate to present our
method since the fast decay of spin correlation functions tends to minimise
the boundary effects on the recursion coefficients. We analyse also the
recursion coefficients corresponding to a recently presented variational
study of the one-hole spectral function using the subspace of one-spinon
wave-functions \cite{Hayn}. It completes our approach by providing
additional information on the asymptotic behaviour of recursion coefficients
and on the spectral function. 
The main physical question connected with the MG model is: how are the
features of spin-charge separation in 1D modified by a spin gap. We would
like to present a rather complete description of the one-hole spectral
function of the MG model. The paper is organised as follows: first we recall
the recursion method (Sec.\ \ref{rmform}) and introduce the one-dimensional 
$t-J-J^{\prime}$ model (Sec.\ \ref{tJJ1}). In Sec.\ \ref{Analyt} we collect
analytical and numerical results in the strong coupling limit $J,J^{\prime
}\to 0$ and in Sec.\  V we present the asymptotic behaviour of the
recursion coefficients. That analysis is used to reconstruct the spectral
function (Sec.\  \ref{SpecDen}) and to calculate the bound states and band
gaps (Sec.\  \ref{BoundSt}).

\section{The Recursion Method}

\label{rmform}

The recursion method (RM) \cite{Haydock72,Haydock80} is a powerful tool to
calculate the matrix elements of the resolvent ($z-\hat H)^{-1}$ of a
Hamiltonian $\hat H$. Given an arbitrary starting state $|u_0\rangle $, the
RM generates a new basis in which the Hamiltonian matrix is tridiagonal. One
can formulate the RM in two ways, in the Hamiltonian or the Liouvillian
representation \cite{VisMul}. The latter one corresponds to a generalisation
of the RM to the Liouvillian space of operators. Usually, it is used for
finite temperature problems. In the present case, however, both formulations
will be demonstrated to be identical and we concentrate on the Hamiltonian
variant.

In the Hamiltonian representation, the RM starts from a vector in Hilbert
space, i.e. from some wave function of the system $|u_0\rangle $. The new
basis (orthogonal, but not normalised) is generated according to the Lanczos
procedure 
\begin{equation}
\label{un}|u_{n+1}\rangle =(\hat H-a_n)|u_n\rangle -b_n^2|u_{n-1}\rangle 
\end{equation}
with $|u_{-1}\rangle =0$, and $b_0^2=\langle u_0|u_0\rangle $. The
coefficients are calculated from 
\begin{equation}
\label{anbn}a_n=\langle u_n|\hat H|u_n\rangle /\langle u_n|u_n\rangle
,\;b_n^2=\langle u_{n+1}|u_{n+1}\rangle /\langle u_n|u_n\rangle \; . 
\end{equation}
Having calculated the recursion coefficients $a_n$ and $b_n$, one can easily
find the matrix element of the resolvent with $|u_0\rangle $: 
\begin{eqnarray}
\label{cf1}
R(z)&=& \langle u_0 | (z - \hat{H})^{-1} | u_0 \rangle  \\
&=& \frac{b_0^2}{
z -a_0-\frac{b_1^2}{z -a_1-\frac{b_2^2}{\ddots }}} \equiv  
\frac{b_0^2}{z-a_0-}\frac{b_1^2}{z-a_1-}\cdots  
\; .
\nonumber
\end{eqnarray}
For a starting state corresponding to one hole in the correlated state, the
resolvent $R(z)$ is identical to the one-particle Green's function and its
imaginary part gives the spectral function we are interested in.

In this form the RM was applied to many body problems (see e.g.\ Refs.\  
\cite{GaglBacc,Dagotto,mb} and references therein) several times. 
In these cases
the resolvent matrix elements of the full many body Hamiltonian were
calculated. This work can be done numerically for finite systems.
Unfortunately, at present, such calculations are only possible for very
small clusters. The resulting $R(z)$ may be related with the dynamic
susceptibility \cite{GaglBacc}. The problem is that $R(z)$ for a finite
system is a rational function that may be represented as a set of poles. The
corresponding spectral density consists of a set of $\delta $-functions
which are usually represented using an artificial broadening. This is quite
different from that what we can expect for a macroscopic system of
interacting particles. The spectrum of a nontrivial interacting system
contains usually a continuous part (the ``continuum'' which is connected
with a branch cut in $R(z)$) and eventually one or several isolated states
(the point spectrum)). A sequence of broadened $\delta $-functions is a very
poor approximation for the continuous part of the spectrum. Furthermore,
such a procedure neglects completely the important difference between the
continuum and isolated poles. All the fine structure of the spectrum is
completely lost.

Below we will show what kind of information about the macroscopic system may
be extracted from the exact diagonalisation studies of small clusters and
propose a more physical procedure of smoothing the spectral density. The ED
has been performed for the 1D $t-J-J^{\prime }$ model with 12, 16, 20, and
24 lattice sites according to the method described in Refs.\ 
\cite{GaglBacc,Dagotto}. From these calculations we extract 
the information about
the bulk related asymptotic behaviour of the recursion coefficients $a_n$
and $b_n$ for large $n$. Additional information about the asymptotics can be
obtained using the variational approach presented in Ref.~\cite{Hayn}.

\section{Majumdar-Ghosh model}

\label{tJJ1} In order to demonstrate our method, we specialise to the
calculation of the one-hole spectral function in the one-dimensional
Majumdar-Ghosh model (the 1D $t-J-J^{\prime }$ model with $J^{\prime}/J=1/2$). 
The interest to 1D models has been revived after the creation of
quasi one dimensional transition metal compounds. The reports on
angle-resolved photoemission spectroscopy (ARPES) \cite{K96} studies show
that the most striking theoretical prediction for 1D strongly correlated
systems, the spin-charge separation, can be observed. The interest to this
phenomenon has increased after Anderson's proposal that it explains the
properties of 2D cuprate superconductors \cite{Anderson}. The calculation of
one-particle Green's functions for 1D strongly correlated system is a complex
and non-trivial task. The complete answer is absent even for the exactly
solvable 1D Hubbard model \cite{LiWu}. That is why the development of simple
approximate approaches remains important. In the present study we consider
the $t-J-J^{\prime }$ model, which describes the basic physics of the
electron motion in 1D transition metal compounds 
\begin{equation}
\label{H}\hat H=\hat t+\hat J+\hat J^{\prime }\;, 
\end{equation}
where 
$$
\hat t=-t\sum_{i,\alpha }\left( X_i^{\alpha 0}X_{i+1}^{0\alpha }+h.c.\right)
\;,
$$
$$
\hat J=\frac J2\sum_{i,\alpha ,\beta }X_i^{\alpha \beta }X_{i+1}^{\beta
\alpha }\;,\ \hat J^{\prime }=\frac{J^{\prime }}2\sum_{i,\alpha ,\beta
}X_i^{\alpha \beta }X_{i+2}^{\beta \alpha }\;, 
$$
and we introduce Hubbard projection operators that act in the subspace of
on-site states, namely boson-like operators 
$$
X_i^{\alpha \beta }\equiv \left| \alpha ,i\right\rangle \left\langle \beta
,i\right| \; , \ \alpha ,\beta =\uparrow ,\downarrow \; , 
$$
and fermion-like ones $X_i^{\alpha 0}$, $i$ being the site index. To make
contact with the standard notation we note that 
$$
\frac 12\sum_{\alpha ,\beta }X_i^{\alpha \beta }X_j^{\beta \alpha }={\bf S}_i
{\bf S}_j+\frac 14n_in_j\;,\ 
$$
for $i\neq j$, where ${\bf S}_i$ and $n_i$ are spin and density operators,
respectively. The exchange term to second neighbours $J^{\prime }$ is
especially relevant in 1D compounds with a spin-Peierls transition, like
CuGeO$_3$. For $J^{\prime }/J$ larger or equal 0.2411 the quasi long-range
order (algebraically decaying spin-correlation functions) of the 1D
antiferromagnetic spin-half Heisenberg model is destroyed and one 
observes a gap in the spin-excitation spectrum \cite{CrtFrst}.

Our aim is to calculate the one-particle two-time retarded Green's function 
\begin{equation}
\label{G}G(k,z)=\langle \langle X_k^{\sigma 0}|X_k^{0\sigma }\rangle \rangle 
\end{equation}
for $z=\omega +\imath 0^{+}$ and the spectral density 
\begin{equation}
\label{A}A(k,\omega )=-\frac 1\pi {\rm Im}G(k,\omega +\imath 0^{+}) \; , 
\end{equation}
where $X_k^{\sigma 0}=\sqrt{2/L}\sum_m{\rm e}^{-\imath km}X_m^{\sigma 0}$,
and $L$ is the number of sites, the factor $\sqrt{2}$ is introduced for
the purpose of normalisation. The above one-particle Green's function 
describes the motion
of one hole in the correlated state described by the Majumdar-Ghosh wave
function \cite{Majumdar} (from now on we consider the special case 
$J^{\prime }/J=1/2$). The notation means 
\begin{equation}
\label{Gfdef}\langle \langle A|B\rangle \rangle \equiv -i\int_{t^{\prime
}}^\infty \!\!dt{\rm e}^{i\omega (t-t^{\prime })}\langle \left\{
A(t),B(t^{\prime })\right\} \rangle \; ,
\end{equation}
with $\left\{ \ldots ,  \right\} $ denoting the anticommutator, and where the
expectation value means the thermal average over a grand canonical ensemble: 
$\langle ...\rangle =Q^{-1}Sp[{\rm e}^{-\beta (H-\mu N)}...],
\hspace{\parindent}Q=Sp{\rm e}^{-\beta (H-\mu N)}$. Here $Sp$ implies the
trace of an operator, $N$ is the particle number operator, $\beta =(kT)^{-1}$
is an inverse temperature, and $\mu $ represents the chemical potential. The
time dependence of the operator $A(t)$ is given by $A(t)={\rm e}^{it(H-\mu
N)}A{\rm e}^{-it(H-\mu N)}$. At zero temperature, $\langle ...\rangle $ goes
over into the expectation value with the ground-state wave function $|\Psi
_0\rangle $. It is not difficult to see that in the given case the
one-particle Green's function (\ref{G}) may be formulated as a resolvent
matrix element 
\begin{equation}
\label{GED}G(k,z)=\langle X_k^{\sigma 0}|(z+E_0-\hat H)^{-1}|X_k^{\sigma
0}\rangle \; , 
\end{equation}
with the state $|X_k^{\sigma 0}\rangle =X_k^{\sigma 0}|\Psi _0\rangle $
denoted in the same way as the corresponding operator. It means also that
the Hamiltonian and Liouvillian formulation of the RM are now equivalent if
we take into account the shift of the diagonal recursion coefficients $a_n$
by the ground-state energy $E_0$ of the pure spin system.

When the ground state is degenerate, as it is the case in the Majumdar-Ghosh
model, we have to take $Sp$ over the ground-state manifold in the
calculation of $\langle ...\rangle $ in (\ref{Gfdef}). We have found that
the spectral function calculated for two orthogonal ground states of the
model according to the prescription given in Sec.\ \ref{SpecDen} coincides 
within the accuracy of the method.

\section{Strong coupling limit}

\label{Analyt}

For the following analysis it is important to recall some former results for
the one-hole spectral function in the MG model using a variational ansatz  
\cite{Hayn}. It was shown that one can give an exact result in the strong
coupling limit $J,J^{\prime }\rightarrow 0$. In this limit only the $\hat t$
operator remains in Eq.\ (\ref{H}). Note that it is a true many-body
Hamiltonian due to the constraint of no double occupancy, it is often called
the $\hat t$-model. Formally, the $\hat t$-model is the limiting case of
various models ($U\rightarrow \infty$ of the Hubbard model, $J\rightarrow 0$
of the $t-J$ model, etc.), and its ground state is strongly degenerate. The
degeneration is removed by an infinitesimal perturbation that fixes the
ground state. It is clear that different spin-models give different answers
for the Green's function (\ref{G}) and the spectral density (\ref{A}) in the
limit of the $\hat t$-model since they may differ by the spin ground state  
\cite{Hayn}.

The exact result was found using the basis operator set 
\begin{equation}
\label{vkr}v_{m,r}=\sqrt{\frac 2L}\sum_{\alpha _1,\ldots ,\alpha
_r}X_m^{\sigma \alpha _1}X_{m+g}^{\alpha _1\alpha _2}\ldots
X_{m+r-g}^{\alpha _{r-1}\alpha _r}X_{m+r}^{\alpha _r0} \; , 
\end{equation}
$g=r/\left| r\right| $ . The operator $v_{m,r}$ may be interpreted as holon
('right end' $\ldots X_{m+r}^{\alpha _r0}$) and spinon ('left end' $
X_m^{\sigma \alpha _1}\ldots $) excitations connected by a string of spin
'flips' (in fact, this string contains also diagonal spin operators but they
also create excitations in a quantum antiferromagnet). The double Fourier
transform over the $m$ and $r$ indices gives exact eigenstates of the 
$\hat t$-term for any magnetic background. In the case of 
finite exchange values, the 
basis set (\ref{vkr}) is not complete and corresponds to the single spinon 
approximation. 

It was shown that the Green's function (\ref{G}) for fixed $k$ has a formal
analogy with the on-site Green's function of the 1D tight-binding model of
non-interacting electrons in a non-orthogonal basis 
\begin{eqnarray}
\label{Gt}&&G(k,z)=\frac 1L\sum_Q\frac{2Z(Q)}{z+\epsilon _h(k-Q)}
\\
&&=\frac 1{
\sqrt{z^2-4t^2}}\left[ 1+\sum_{n=1}^\infty 2\cos kn\left\langle \Omega
_{0\rightarrow n}\right\rangle \tau ^n(z)\right] , 
\nonumber
\end{eqnarray}
where $\epsilon _h(k)\equiv 2t\cos (k)$. 

The spectral function may be represented like 
\begin{equation}
\label{ar}A(k,\omega )=\int_{-\pi }^\pi \frac{dQ}\pi Z(Q)\delta (\omega
+\epsilon _h(k-Q)) \; , 
\end{equation}
with 
\begin{eqnarray}
&&Z(Q)\equiv \frac 12\sum_{n=-\infty }^\infty \exp \left[ -\imath (Q-\pi
)n\right] \left\langle \Omega _{0\rightarrow n}\right\rangle \; , 
\\
&&\Omega _{0\rightarrow r}=\sum_{\alpha _1,\ldots ,\alpha _r,\sigma
}X_0^{\sigma \alpha _1}\ldots X_{r-g}^{\alpha _{r-1}\alpha _r}X_r^{\alpha
_r\sigma }
\nonumber \\
&&=(2{\bf S}_0{\bf S}_g+\frac 12)(2{\bf S}_g{\bf S}_{2g}+\frac 12
)\ldots (2{\bf S}_{r-g}{\bf S}_r+\frac 12) \; , 
\nonumber \\
&&\label{tau}\tau (z)\equiv \frac{z-\sqrt{z^2-4t^2}}{2t}=\frac t{z-}\frac{t^2}{
z-}\frac{t^2}{z-}\cdots \; . 
\nonumber
\end{eqnarray}
Integrating over the $\delta $-function, the spectral function (\ref{ar})
may be given in the form \cite{SP} 
\begin{equation}
\label{Aansw}A(k,\omega )=\frac{Z(k+\phi )+Z(k-\phi )}{\pi \sqrt{4t^2-\omega
^2}}\;,\qquad \cos (\phi )=-\frac \omega {2t}\;.  
\nonumber
\end{equation}
The expressions (\ref{ar},\ref{Aansw}) imply that the spectrum is 
continuous and bounded on the interval $-2t\leq \omega \leq 2t$ for all the
possible spin models in the considered limit. It may be interpreted as holon
motion with the dispersion $\epsilon _h(k)$ with an immovable spinon. The
difference in hole spectral densities and Green's functions arises due to
the expectation value $Z(Q)$ that depends on the spin part of the
Hamiltonian. Some parts of the spectrum may be excluded by a vanishing $Z(Q)$.

For the 1D pure (non frustrated) antiferromagnetic Heisenberg model, $Z(Q)$
was given in Ref.\ \cite{OgSh,SP} as 
$Z(Q)\newline \propto \Theta (Q-\pi /2)/\sqrt{Q-\pi /2}$. The square-root 
singularity leads to additional peaks in the
spectral function, such that for each $k$ value one can distinguish well
pronounced spinon and holon peaks (for details see also Ref.\ 
\cite{Nagaosa97}).
For the Majumdar-Ghosh (MG) wave function, on the other hand, one finds $
Z(Q)=\frac 32\left( 1+\cos Q\right) /\left( 5+4\cos 2Q\right)$ 
(see Ref.\ \cite{Maekawa}) where $Z(Q)$ is non zero for all 
values of $Q$. The peak at $Q=\pi
/2$ becomes very wide. That leads to a large damping of the holon 
peak \cite{Hayn}. One finds the following explicit formula for the 
spectral function
of the MG model in the limit $J,J^{\prime }\to 0$: 
\begin{eqnarray}
\label{AMG}&&A(k,\omega )=\frac 3{8t\pi \sqrt{1-x^2}}
\cdot
\nonumber \\
&&
\frac{20+4x\cos
k+16(2x^2-1)\cos 2k-8D}{25+40(2x^2-1)\cos 2k+8\cos
4k+8(8x^4-8x^2+1)}, 
\nonumber \\
&&
x\equiv \frac \omega{2t} \; ,  \quad D\equiv x\cos 3k+4x^3\cos k
\end{eqnarray}
which is non zero for all $-2t\leq \omega \leq 2t$.

Now, for the MG model, we compare the continued fraction expansion of the
one-particle Green's function for the macroscopic system (\ref{Gt}) with
that for a finite ring. The summation of series in (\ref{Gt}) gives 
\begin{equation}
\label{Gmg}G(k,z)=\frac{4-2\left[ 2\tau (z)+\tau ^3(z)\right] \cos k-\tau
^4(z)}{\sqrt{z^2-4t^2}\left[ 4+4\cos 2k\tau ^2(z)+\tau ^4(z)\right] } \; ,
\end{equation}
that may be rewritten using the identity $\tau (z)=t/(z-t\tau (z))$ in the
form 
\begin{equation}
\label{Gmgcf}G(k,z)=\frac 1{z-a_0-}\frac{b_1^2}{z-a_1-b_2^2t^{-1}\tau
(z)\left[ 1+F(z)\right] } \; ,
\end{equation}
where 
\begin{eqnarray}
\label{abmg}&& a_0=\frac t2\epsilon \, ,\ b_1^2=3t^2(1-\frac 34\epsilon ^2) 
\, , \ a_1=\frac{9\epsilon ^3-6\epsilon }{8b_1^2}t^3 \; , 
\nonumber \\
&&   
b_2^2=\frac{9(8-7\epsilon ^2)t^4}{16b_1^2} \; , \ \epsilon \equiv \cos k \, , 
\end{eqnarray}
and the function $F(z)$: 
\begin{eqnarray}
F(z)&=&\frac{\epsilon (1-\epsilon ^2) \ V(z) }{(8-7\epsilon ^2) \ W(z)} \; ,
\nonumber \\
&& V(z) \equiv 
z(18\epsilon ^4-27\epsilon
^2+8)+4\epsilon (10-9\epsilon ^2)
\nonumber \\
&& - 3(12\epsilon ^4-21\epsilon ^2-24)
\tau(z) \; ,
\nonumber \\
&& W(z) \equiv  
2z^2(4-3\epsilon ^2)-2z\epsilon
-12
\nonumber \\
&& + 11\epsilon ^2-\left[ 2z(4-3\epsilon ^2)-\epsilon \right] 
\tau(z)\; ,
\nonumber
\end{eqnarray}
vanishes for $k=0,\pm \frac \pi 2,\pm \pi $. For these $k$ values we thus
readily obtain $G(k,z)$ in the continued fraction form (\ref{cf1}) with $
a_n=a_\infty =0,\ b_n^2=b_\infty ^2=t^2,\ n>2$ that follows from (\ref{tau}). 
For other $k$ values the coefficients asymptotically tend to the same
constant limits $a_\infty ,b_\infty $.
\begin{figure}
\resizebox{0.45\textwidth}{!}{  \includegraphics{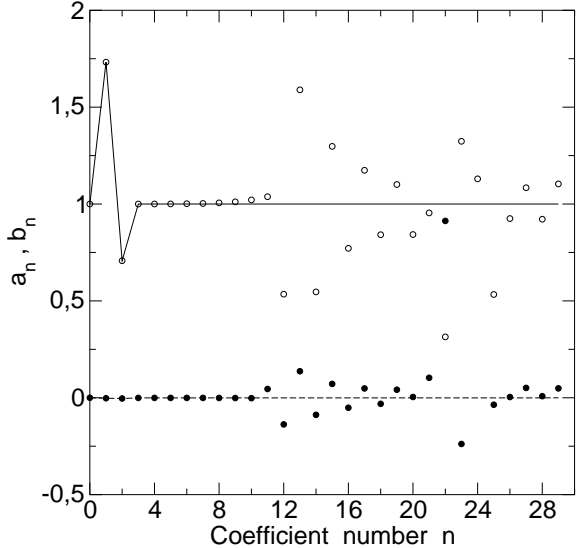}}
\caption{
The coefficients $a_n(k)$ (filled circles) 
and $b_n(k)$ (open circles) of the continued fraction expansion
of the one-particle Green's function $G( k,\omega )$ versus $n$  for 
$k=\pi /2$, $J=0.01$ provided by the exact diagonalisation 
of the $L=24$ sites $t-J-J'$ ring  compared with the result for 
the single spinon approximation (solid or dashed line, exact for 
$J=0, L=\infty $). 
The hopping parameter $t=1$ is the unit of energy.  
}
\end{figure}

Fig.~1 compares the analytical result for $a_n,b_n$ with the values
provided by the ED of a ring with $L=24$ sites at $J=0.01$. One sees that
the values of $a_n$ with $n\leq 10$ are not influenced by the boundary.
These first values give us bulk-related information. Comparing with other
cluster sizes $L$, one finds that $a_n$ has no finite size effect for $n\leq
(L-4)/2$. For the off-diagonal elements $b_n$ one has $n\leq (L-4)/2+1$. So,
we may conclude that ED studies of finite systems are able to provide the
information that concerns the macroscopic system and may be used for the
analysis of the spectral density. Despite the small number of 'bulk related'
coefficients they definitely reached their asymptotic regime and may be
extrapolated to $n\rightarrow \infty $ giving the {\em exact} Green's
function and spectral density for infinite chain. The situation is not
always so simple, but this example clearly shows a manageable way to
extrapolate ED results to the macroscopic system.

Concluding the consideration of the strong coupling limit we recall that it
is commonly believed that the main changes in the spectral density at
larger finite couplings (i.e. $J\neq 0$) come from the appearance of the
spinon dispersion $\epsilon _s$, and may be accounted for by the
substitution 
\begin{equation}
\label{es}\epsilon _h(k-Q)\rightarrow \epsilon _h(k-Q)+\epsilon _s(Q) 
\end{equation}
in Eqs.(\ref{Gt},\ref{ar}) \cite{SP,Nagaosa97}. Below we shall
demonstrate that holon and spinon scattering lead to additional changes:
the continuous part of the spectrum expands up to infinite energies and for
the case of the MG model, additional discrete states appear.

\section{Termination of the continued fraction}

\label{term} Now we concentrate on results for finite $J,J^{\prime }$. In
the previous section we have shown that the few first recursion coefficients
given by the exact diagonalisation study of a small cluster are (almost) not
affected by boundaries. Thus, in order to infer the shape of the spectral
density for macroscopic system we should rely on these 'bulk related'
coefficients. 
To define the bulk related coefficients we observe from Fig.\ 2a that the
coefficients for $L=12$ start to deviate from the common line at $n=4$,
those for $L=16$ at $n=6$ and for $L=20$ at $n=8$. The corresponding numbers
for $b_n$ (Fig.\ 2b) are 5, 7, and 9. Generalising we find $a_n:n<(L-4)/2$
and $b_n:n<(L-4)/2+1$ to be bulk related coefficients. From the inset of
Fig.\ 2a  
we clearly see that 'bulk related' coefficients for finite $J,J^{\prime }$
show a different tendency (to grow with moderate oscillations) compared to
the rest of the sequence for a cluster (strong oscillations around a
constant value). 
\begin{figure*}
\resizebox{0.45\textwidth}{!}{  \includegraphics{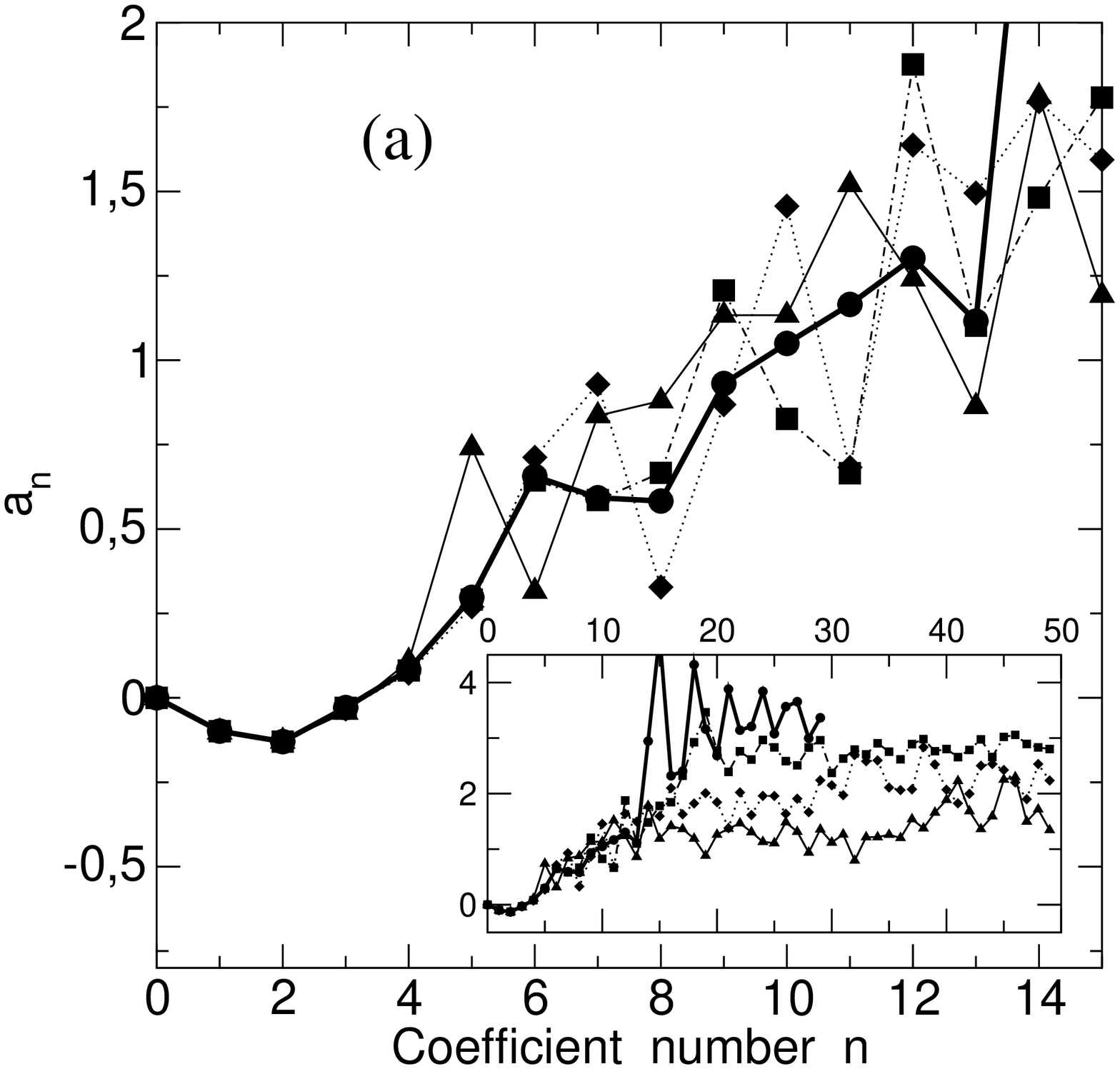} }
\resizebox{0.45\textwidth}{!}{  \includegraphics{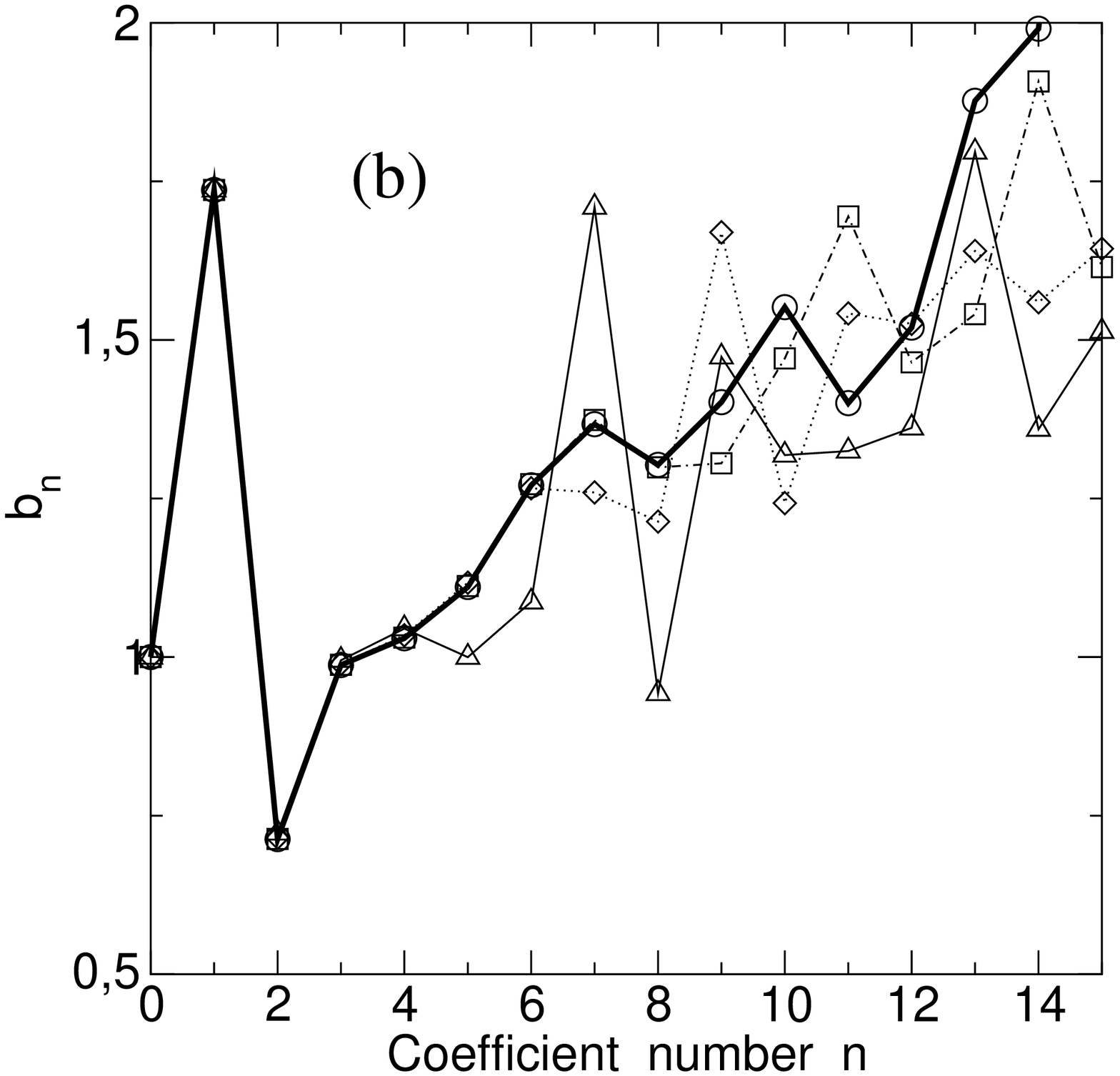} }
\caption{
The coefficients $a_n(k)$ (full symbols, (a)), and 
$b_n(k)$ (open symbols, (b)) of the continued fraction
expansion  of the one-particle Green's function $G(k,\omega )$ versus $n$ for 
$k=\pi /2$, $J=0.4$ and various lattice sizes $L=12$ 
(triangles and thin solid line), $16$ (diamonds and dotted line), 
$20$ (squares and dash-dotted line), $24$ (circles and thick solid line). 
The insert in panel (a) shows $a_n(k)$ on a larger scale. 
}
\end{figure*}

For finite $J$ the number of bulk related coefficients is
slightly reduced in comparison to Fig.~1. It should be noted that
the 'bulk related' coefficients with large $n$ have still a small finite
size effect, but that is several orders of magnitude smaller than for the
rest of the sequence. We have also observed that in the case of the pure
$t-J$ model (i.e.\ $J^{\prime}=0$) the situation is worse. The number 
of recursion
coefficients that are not influenced by finite size effects is considerably
reduced. One may recall that according to Eq.\ (\ref{anbn}) the coefficients
are related to static spin correlation functions of the spin background and
speculate that the difference arises due to the algebraically decaying
spin correlation functions for the 1D Heisenberg model in difference to the
exponential decay for the MG model.

For the calculation of the spectral density we will try to extrapolate the
tendency shown by the 'bulk related' coefficients. In other words, we
rewrite the expression (\ref{cf1}) in the form 
\begin{equation}
\label{cf2}G(k,z)=\frac{b_0^2}{z-a_0-}\frac{b_1^2}{z-a_1-}\cdots \frac{
b_{n_0}^2}{z-a_{n_0}-T_{n_0}(k,z)} \; , 
\end{equation}
and try to find a function $\tilde T_{n_0}$ (a so called ''terminator'') that
is close to $T_{n_0}$. The terminator should have such an analytic behaviour
which corresponds to the asymptotics of $a_n$ and $b_n$.

Various ways to construct such approximations are described in the
literature on the recursion method (see Refs.\ 
\cite{HayNex85,Ducastelle,Nex78,Magnus85}). The asymptotic 
behaviour of continued
fraction coefficients is governed by the band structure and the singularities 
of the spectral density \cite{Magnus85}. The problem is well studied for a 
bounded spectrum: $\{a_n\}$ and $\{b_n\}$ converge toward limits in the
single band case, oscillate endlessly in a predictable way in the multiband
case. Damped oscillations are created by isolated singularities. But the
growth of coefficients indicates that the spectrum we deal with is
unbounded. The infinite growth of diagonal coefficients is observed for the 
$t-J_z$ model that differs from the $t-J$ model by omitting spin
fluctuations, $a_n=nJ_z,\ b_n^2=z_0t^2$, ($z_0$ is the number of nearest
neighbours) \cite{Izing96}. This growth has a clear physical meaning: a hole
moving on the Neel background (the ground state for $t-J_z$ model without
holes) creates a string of overturned spins; every $|u_n\rangle$ state 
(\ref{un}) is
the combination of such strings containing $n$ overturned spins, thus its
energy $\left\langle u_n\right| \hat t+\hat J_z\left| u_n\right\rangle
\propto nJ_z$. The non-diagonal coefficients $b_n$ are constant. As a result,
the hole is localised in the $t-J_z$ model, its spectral function does not
depend on quasimomentum $k$ and consists of an infinite set of $\delta $
-functions. The spectrum is discrete and unbounded \cite{IzNote}. 

When spin fluctuations are added the situation becomes more complicated, now
the non-diagonal coefficients also grow. Such a situation was studied in the
two-dimen\-sional $t-J$ model as well as for spin-fermion models of the CuO$_2$
plane, where the following asymptotics was found within the framework of
the self-consistent Born approximation \cite{SCBA,linterm}:  
\begin{equation}
\label{asab}b_n\approx \lambda _1n+\lambda _2 \, ,\quad 
a_n\approx 2\lambda_1n+\lambda _3 \, ,\quad 
\lambda _i=\lambda _i(k),\quad n\gg 1 \; . 
\end{equation}
It indicates the existence of an exponential tail in the spectral density.
But the connection between an unbounded spectrum (toward high energies) and
the growth of recursion coefficients is far more general. For example, the
model spectral density 
$$
A(\omega )=c^t\omega ^\rho \exp (-\omega ^\beta ),\ c(\beta )=\frac{2\Gamma
(\beta )}{\left[ \Gamma (\beta /2)\right] ^2},\ \omega >0 \; ,
$$
with the parameters $t$, $\rho$, and $\beta$,
gives the following asymptotic behaviour of the recursion coefficients \cite
{Magnus85}: 
$$
b_n\approx \left( \frac n{c(2\beta )}\right) ^{1/\beta },\quad a_n\approx
2\left( \frac n{c(2\beta )}\right) ^{1/\beta } \; ,
$$
which does not depend on $t$ and $\rho$. 
For $\beta =1$ we have a linear dependence on $n$ for $\{a_n\}$ and 
$\{b_n\}$ \cite{Haydock80}. 

The linear growth holds also for the MG model. Fig.~3 shows the linear
contributions to the growth of 'bulk related' coefficients. We have found $
\lambda _1,\lambda _2$ using least mean square fit of the $\{b_n\}$ 
sequence and only $\lambda _3$ was adjusted for $\{a_n\}$. 
Fig.\ 3 shows that
our assumption about the slope $da_n/dn\approx 2db_n/dn$ holds.
\begin{figure*}
\resizebox{0.45\textwidth}{!}{  \includegraphics{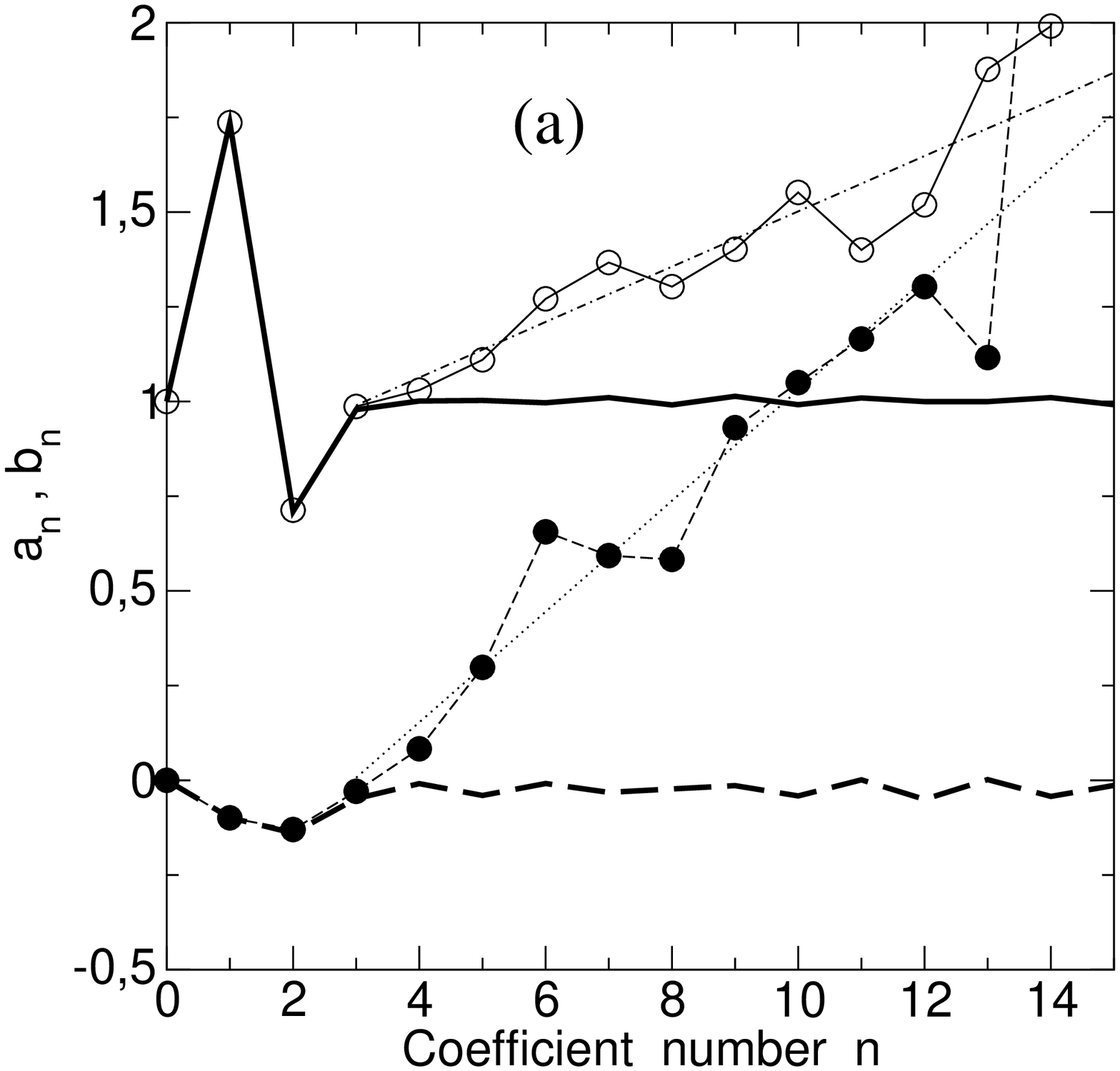} }
\resizebox{0.45\textwidth}{!}{  \includegraphics{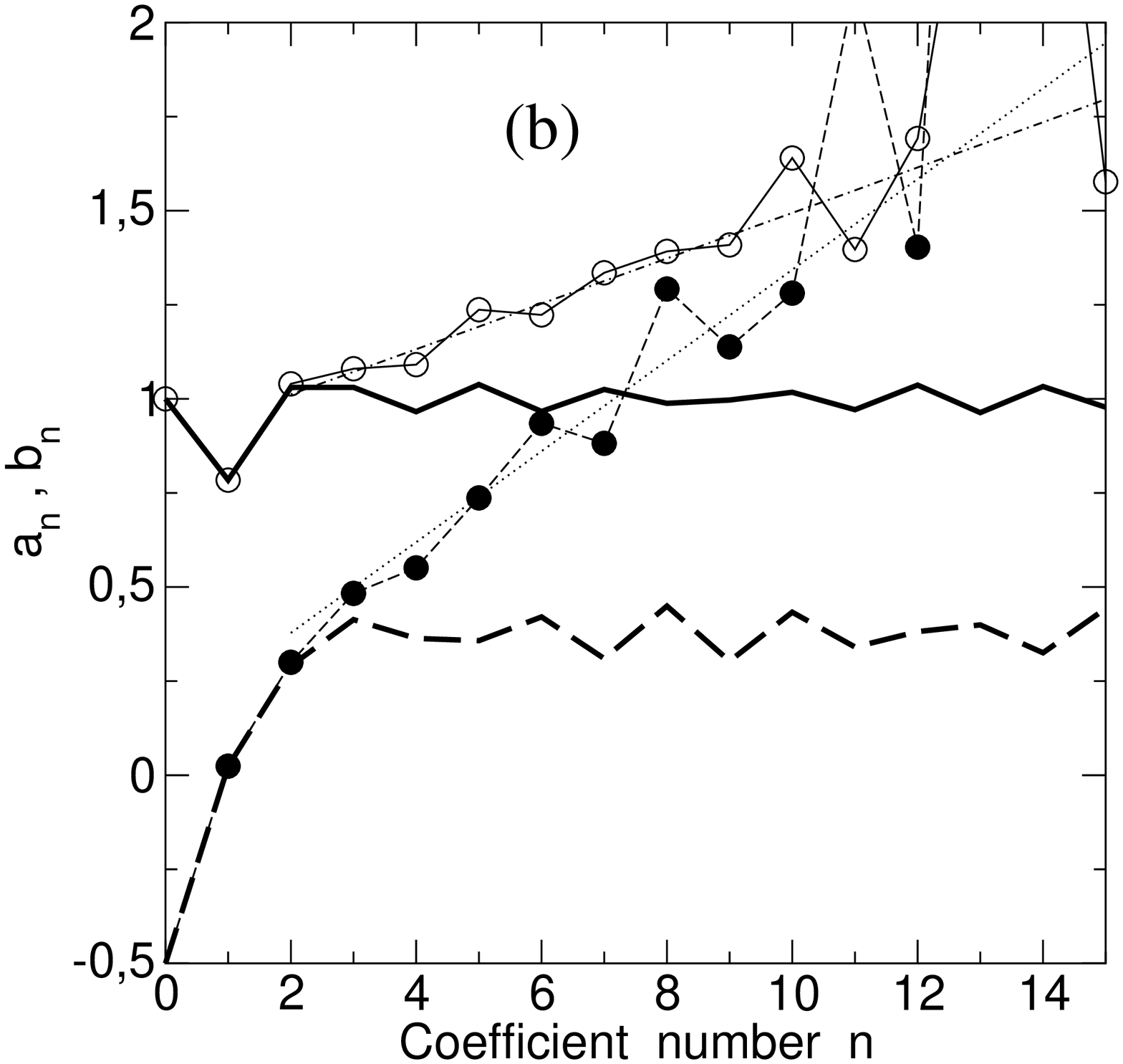} }
\caption{
Comparison of the coefficients $a_n(k)$ (filled circles joined by
dashed lines) 
and $b_n(k)$ (open circles, solid lines) provided by the exact diagonalisation 
of the $L=24$ sites $t-J-J'$ ring  with the result of 
the single spinon approximation (thick dashed and solid lines, 
correspondingly)
for finite $J=0.4$. The quasimomenta are $k=\pi /2$  (a) and $ k=0$ (b). 
The linear contributions in the asymptotics of the 
exact coefficients ($n \rightarrow \infty $:  
$\tilde{b}_n=c_1n+c_2$, $\tilde{a}_n=2c_1n+c_3$) 
are shown by dashed-dotted and dotted lines, correspondingly.
}
\end{figure*}

The continued fraction expansion of the incomplete Gamma function  has the 
same asymptotics (\ref{asab}) and is written as \cite{NR} 
\begin{eqnarray}
\label{gami1}&&\Gamma (\alpha ,x)=\frac{{\rm e}^{-x}x^\alpha }{x+1-\alpha -} 
\frac{1\cdot (1-\alpha )}{x+3-\alpha -}\frac{2\cdot (2-\alpha )}{x+5-\alpha
- }\cdots 
\nonumber \\
&& \cdots
\frac{n\cdot (n-\alpha )}{x+(2n+1-\alpha )-\cdots } \; . 
\end{eqnarray}
We shall use this circumstance for the construction of the terminator 
$\tilde T_N(k,z)$ for $G(k,z)$ (\ref{cf2}). 

Let us introduce the model Green's function 
\begin{equation}
\label{gami2}\tilde G(k,z)=-\frac{\Gamma (\alpha ,x{\rm e}^{-\imath \pi })
{\rm e}^{\imath \pi \alpha }}{\lambda _1{\rm e}^xx^\alpha } \; ,
\end{equation}
with the spectral density 
\begin{equation}
\label{Awave}\tilde A(k,\omega )=\frac 1{\lambda _1\Gamma (1-\alpha )}\left[ 
\frac{(\omega -\omega _0)}{\lambda _1}\right] ^{-\alpha }\exp \left[ -\frac{
(\omega -\omega _0)}{\lambda _1}\right] \; . 
\end{equation}
Here $x=(z-\omega _0)/\lambda _1$, 
\begin{equation}
\label{om0}\omega _0\equiv \lambda _3-2\lambda _2-\lambda _1 \; ,
\end{equation}
is the lower bound of the continuous spectrum, 
$\alpha =2(M-\lambda _2/\lambda _1)$, $M$ is the integer part of 
$\lambda _2/\lambda _1$. The continued
fraction expansion coefficients for $\tilde G(k,z)$ are 
\begin{eqnarray}
\label{gami7}&&\tilde a_m=\lambda _1(2m+1-\alpha )+\omega _0,\ m\geq 0,\ 
\tilde b_0=1 \; , 
\\
&&\tilde b_m=\lambda _1\sqrt{m(m-\alpha )}\approx \lambda
_1(m-\alpha /2),\ m>0.
\nonumber
\end{eqnarray}
Comparing with the definition (\ref{asab}) we find 
$$
b_n\approx \tilde b_{M+n} \, ,\ a_n\approx \tilde a_{M+n} \; . 
$$
Then we slightly generalise the recipe of Ref.\ \cite{HayNex85} and
approximate \cite{linterm} 
\begin{eqnarray}
\label{Twave}\tilde T_{M+n_0}(k,z)&=&\frac{\tilde q_{M+n_0-1}(k,z)-\tilde G
(k,z)\tilde p_{M+n_0}(k,z)}{\tilde q_{M+n_0-2}(k,z)-\tilde G(k,z)\tilde p
_{M+n_0-1}(k,z)}
\nonumber \\
&\approx& T_{n_0}(k,z) \; .
\end{eqnarray}
Here $\tilde p_n(k,z),\tilde q_n(k,z)$ are polynomials of first and second
kind orthogonal with respect to the spectral density (\ref{Awave}). They are
calculated by the recursion relation \cite{Haydock80,Nex84} 
\begin{eqnarray}
&&\tilde p_{-1}=\tilde q_{-1}=0 \; ,\quad 
\tilde p_0=1,\ \tilde q_0=\tilde b_0^2 \; , 
\\
&&\tilde p_n=(z-\tilde a_{n-1})\tilde p_{n-1}-\tilde b_{n-1}^2\tilde p_{n-2}
\; , 
\nonumber \\
&&\tilde q_{n-1}=(z-\tilde a_{n-1})\tilde q_{n-2}
-\tilde b_{n-1}^2\tilde q_{n-3} \; .  
\nonumber
\end{eqnarray}
$\tilde p_n(k,z)$ are proportional to the generalised Laguerre polynomials.
In our calculations it holds always: 
$\lambda _2/\lambda _1>0$ and $-1<\alpha <0$.

We thus obtain $G(k,z)$ in the whole complex energy plane. It has the correct
analytic properties and coincides with the retarded Green's function for $z$
in the upper half plane and with the advanced Green's function for $z$ in 
the lower half plane. It has a branch cut on the real axis for 
$\omega _0<z<\infty $ that
corresponds to the continuum part of the spectrum. In addition, it may have 
isolated poles for $z<\omega _0$ which represent quasiparticle excitations.

Figs.\ 2 and 3 show that the recursion coefficients oscillate
around their asymptotics. The influence of these oscillations on the 
spectral density was neglected so far. For a bounded spectrum this is the 
indication of a 
multiband spectrum. The problem was well studied in Refs.\ 
\cite{Ducastelle,Magnus85}. In the case of one gap, i.e.\ a 
spectrum lying in the disjoint
intervals $E_1\leq \omega \leq E_2$ and $E_3\leq \omega \leq E_4$ ($E_2<E_3$) 
any pair of neighbouring coefficients $(y,x)=(b_n^2,a_n)$ or $
(b_{n+1}^2,a_n)$ obeys the law \cite{Ducastelle} 
\begin{equation}
\label{recTDT}\left( x^2+A_1x+A_2+2y\right) ^2=X(-A_1-x) \; , 
\end{equation}
with
\begin{eqnarray}
&&A_1\equiv \frac 12\sum_iE_i \, , \ 
A_2\equiv \frac 12\sum_{i<j}E_iE_j-\frac{A_1^2}2 \, ,  
\nonumber \\
&& X(z)\equiv \prod_i\left( z-E_i\right) \; . 
\nonumber
\end{eqnarray}
This relation may be represented in phase space. In the asymptotic limit,
all pairs $(b_n^2,a_n)$ and $(b_{n+1}^2,a_n)$ should lie on a single closed
curve (\ref{recTDT}). The best fit provides the band edges directly. After
establishing the $E_i$ values, the relation (\ref{recTDT}) may be used for
the extrapolation of coefficients beyond the known values. Fig.~4 
shows the plot for the single spinon approximation (which leads to a bounded
spectrum) at $k=0$ and $J=0.4$ . Several known values and their
extrapolation form a single curve. An analogous analysis for different $k$
and $J$ has shown that in the single spinon approximation we have one gap in
the continuous part of the spectrum. The asymptotic regime is reached for 
$n>n_0\sim 20$ for $k=0,\pi$. For the more structured spectral density at 
other $k$ values the asymptotics is reached after $n_0\sim 70$. 
The bounds of the
coefficients are 
$$
\frac{E_1+E_4}2-g<a_n<\frac{E_1+E_4}2+g \, , \ g\equiv \frac{E_3-E_2}2 \; , 
$$
$$
\frac{E_4-E_1}4-\frac g2\leq b_n\leq \frac{E_4-E_1}4+\frac g2 \; , 
$$
and the half gap value $g\approx 0.074$ equals the amplitude of
oscillations. This relation between the amplitude of the coefficient 
oscillations and the gap does not depend on the total bandwidth $E_4-E_1$ .
One may expect that it holds also for an unbounded spectrum. We have modelled
such a spectrum by the creation of a gap in the spectral density 
(\ref{Awave}). Then we have calculated the coefficients from the gapped 
spectral density and we have 
found that they oscillate around their asymptotics (\ref{asab}) with an
amplitude equal to half of the gap. As we have mentioned above, the internal
singularities lead to damped oscillations of the coefficients. The restricted
number of exact 'bulk-related' ED coefficients makes it impossible at
present time to analyse their oscillations in detail and to extract
quantitatively the values and positions of the gaps.
\begin{figure}
\resizebox{0.45\textwidth}{!}{  \includegraphics{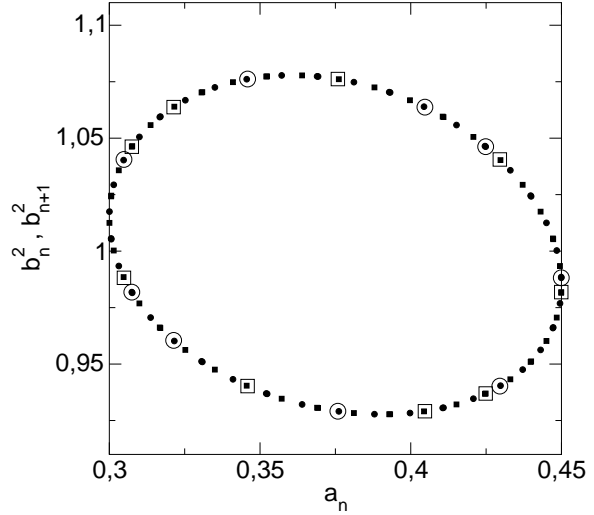} }
\caption{
Phase space representation of the recurrence law (\ref{recTDT}).
The open circles (squares) are pairs of $\{b_n^2, a_n \}$ 
($\{b_{n+1}^2 ,a_n \}$), the full symbols 
correspond to extrapolated values.
}
\end{figure}

\section{Reconstruction of spectral density}

\label{SpecDen}

One may summarise the following recipe of spectral density reconstruction
from the Lanczos recursion coefficients: i) Establish the number of recursion
coefficients that are not strongly affected by the boundaries. ii) Investigate
their asymptotics and find a model function that has a continued fraction
expansion with similar asymptotics, or equivalently find a reliable procedure
to extrapolate the coefficients. iii) Calculate the Green's function $
G(k,z) $ using expressions (\ref{cf2},\ref{Twave}) (then $z$ may be real)
or directly (\ref{cf1}) (with complex $z$, the numerical efforts 
depend on the distance to the real axis) up to convergence. Then the spectral
density results from the definition (\ref{A}).

As we have already mentioned, the number of bulk-related coefficients for
the Majumdar-Ghosh model (\ref{H}) is $n_0\approx (L-4)/2$, where $L$ is the
number of sites. We have performed the ED for rings of the length up to 24
sites, i.e. we have about 10 pairs of coefficients for every $k$ value. The
last 6 pairs were used for the determination of the parameters of the linear
law (\ref{asab}) which are necessary to terminate the continued fraction.
Fig.~5 shows the spectral density $A(k,z),\ z=\omega +\imath \eta 
$ for $k=\pi /2$, $J=0.4$, $\eta =0.01$. On panel (a) the spectral density
calculated according to the above recipe (solid line) is compared with the
spectral density for the $L=24$ sites ring (dotted line) calculated with the
same broadening $\eta $. The qualitative difference of both curves is
evident. Panel (b) of the figure shows the result of the single spinon
approximation. With a broadening of $\eta =0.01$ the overall shape of the
curve is similar to the ED result and is characterised by two peaks at $
\omega \approx \pm 2t$. The small feature near $\omega \sim 0.4$ corresponds
to a bound state in the gap, but all bound states and gaps are smeared out
for a broadening of $\eta =0.01$ (compare with Fig.\ 6b). The essential
difference with respect to the ED spectrum (for the single spinon approximation
the spectrum is bounded) manifests itself as a difference of the relative
weight of the peaks, the higher energy peak of ED being smoothed by the 
exponential tail of the spectral density that tends up to $\omega
\rightarrow \infty $.
\begin{figure*}
\resizebox{0.45\textwidth}{!}{  \includegraphics{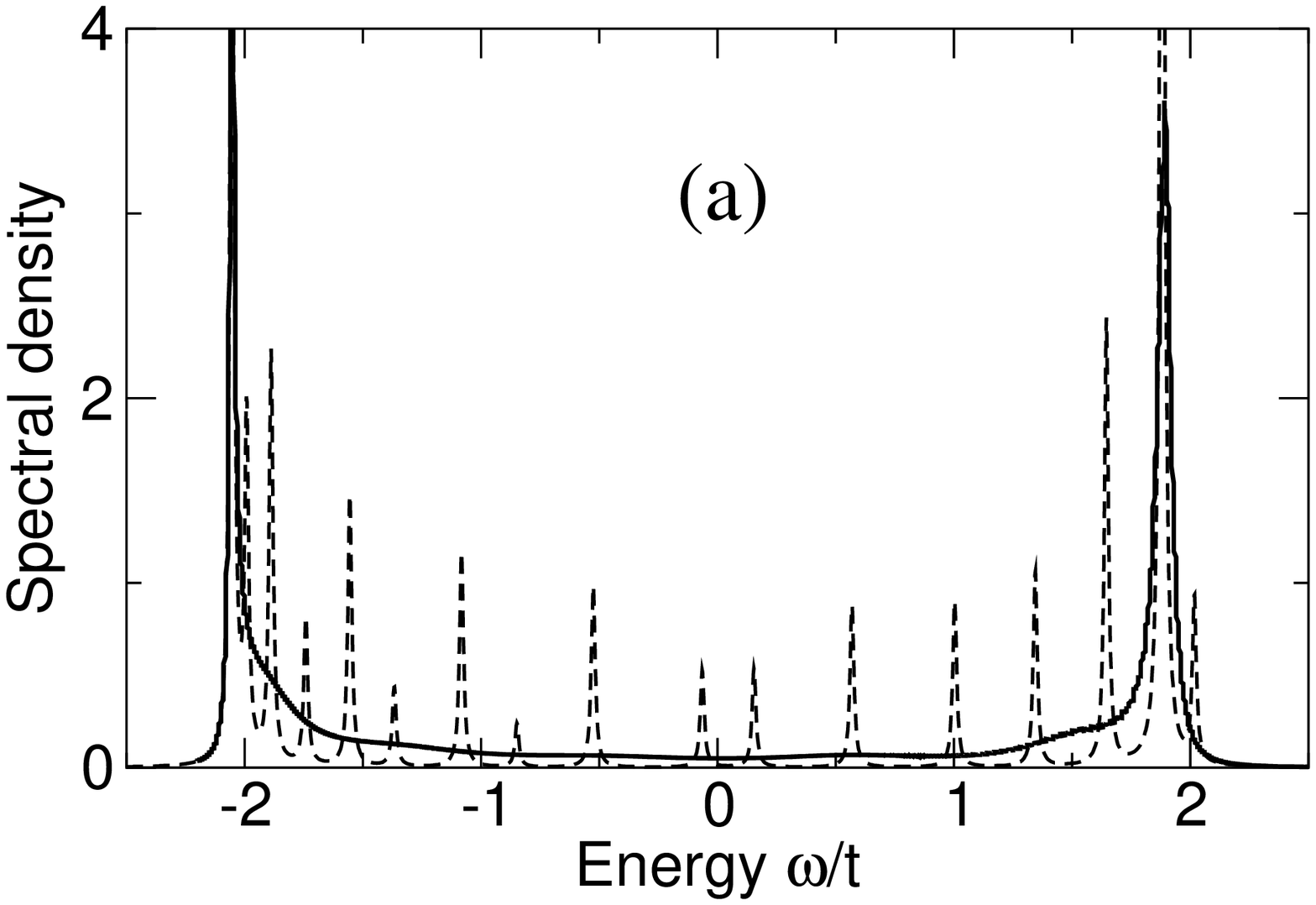} }
\resizebox{0.45\textwidth}{!}{  \includegraphics{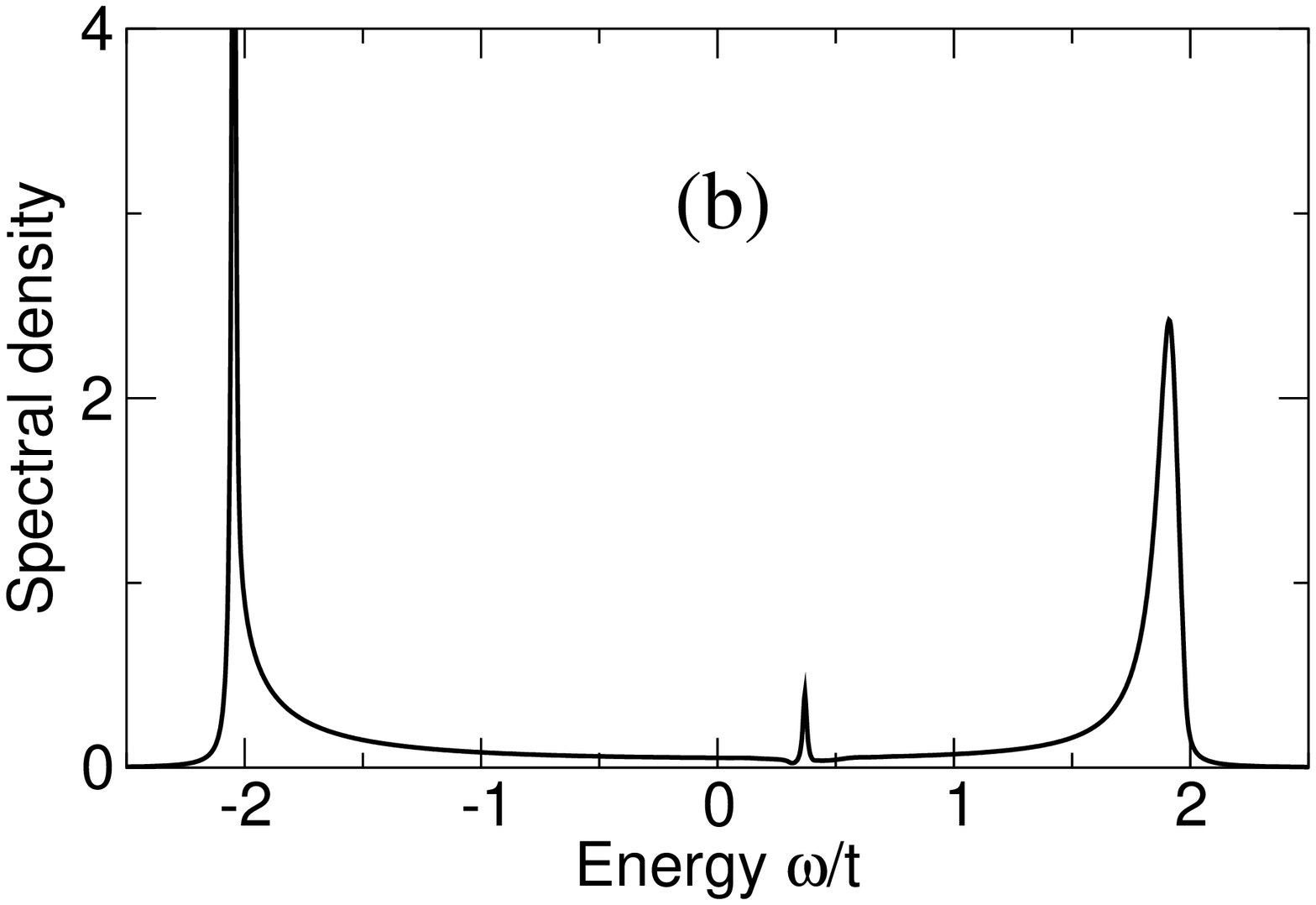} }
\caption{
The spectral density $A(k,\omega +\imath \eta )$ 
for $k=\pi /2$, $J=0.4$, $\eta = 0.01$ (a), compared with the 
single spinon approximation (b), also for $\eta=0.01$. The solid line in panel 
(a) corresponds to the
analytic termination of
the exact Green's function continued fraction expansion  after 9
levels. The termination is based on the coincidence of the asymptotic
behaviour of the continued fraction expansion of the incomplete Gamma
function with the asymptotics of the coefficients shown in Figs.\ 2 and 
3a. 
The  dashed line is the spectral density for the ring with $L=24$ sites.
}
\end{figure*}

Fig.~6 shows the spectral density $A(k,\omega +\imath \eta )$ for
various $k$-values, $J=0.4$. Panel (a) corresponds to the ED with $\eta =0.01
$. The interpretation of the various peaks as different collective
excitations were already broadly discussed in Ref.\ \cite{Hayn}: 
the lower edge of
the continuum corresponds to the spinon excitation with a small dispersion
of the order $J$. For $k$ different from $\pi/2$ one can also observe the
holon peak with a much larger dispersion (proportional $t$) and with a large
damping  \cite{Hayn}. 
\begin{figure*}
\resizebox{0.45\textwidth}{!}{  \includegraphics{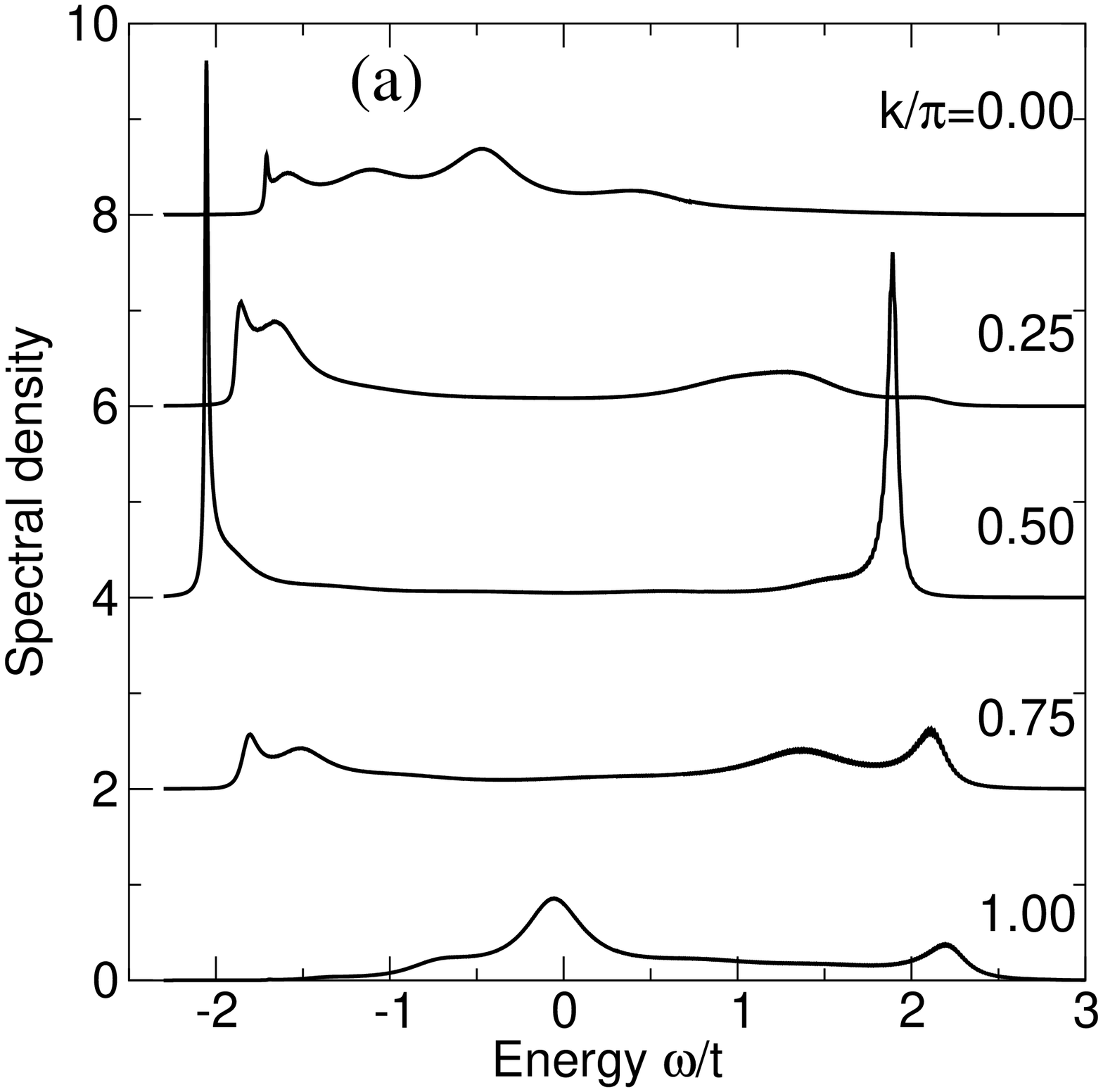} }
\resizebox{0.45\textwidth}{!}{  \includegraphics{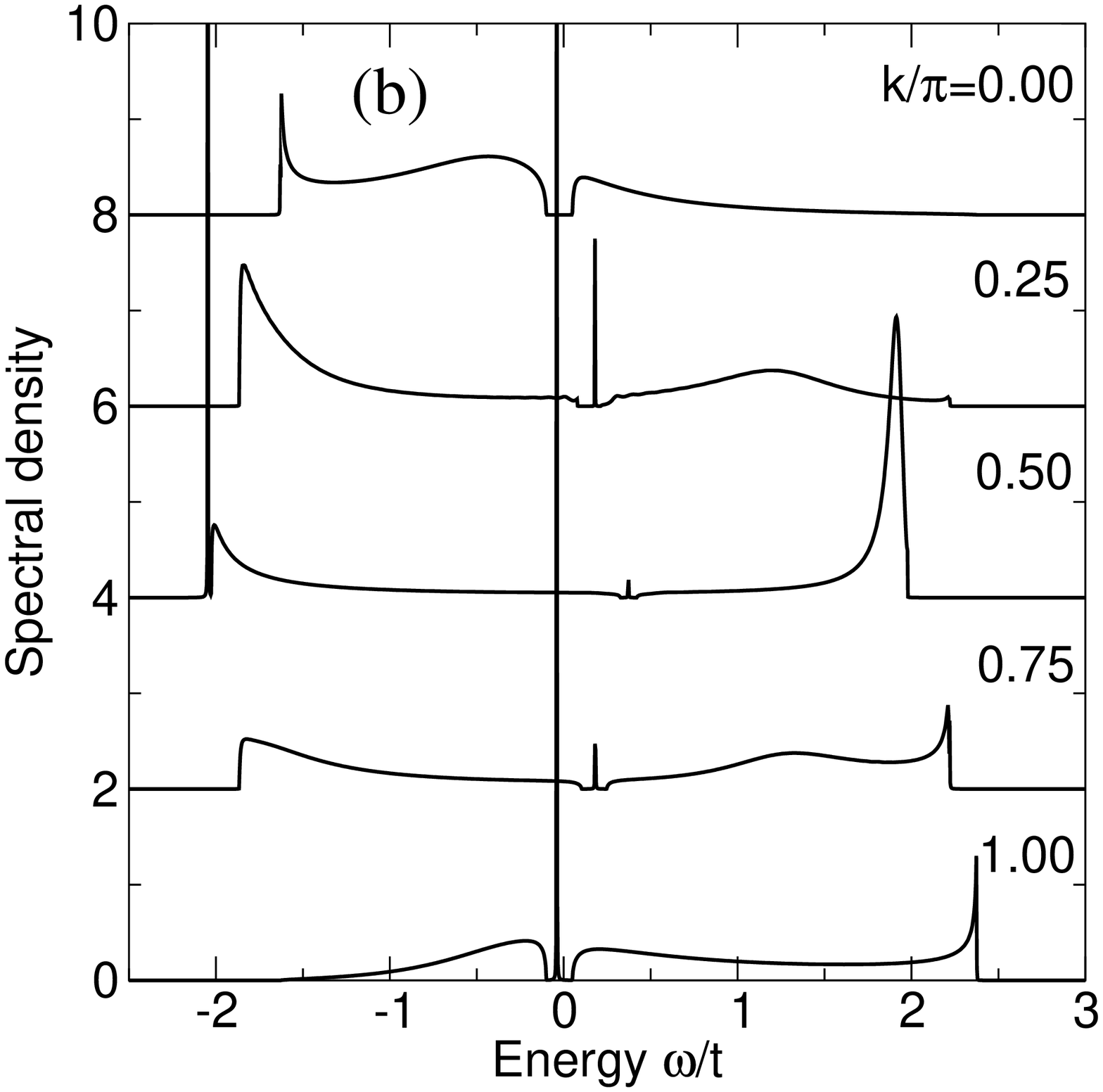} }
\caption{
The spectral density $A(k,\omega +\imath \eta )$  
restored from the 'bulk related' coefficients 
of the exact Green's function continued fraction expansion 
for various $k$-values, $J=0.4$, $\eta = 0.01$ (a).
The spectral density $A(k,\omega +\imath \eta )$ in the single spinon 
approximation (b) 
for various $k$-values, $J=0.4$, with $\eta = 0.0001$.  
}
\end{figure*}

For the single spinon approximation the law (\ref{recTDT}) makes it possible
to reveal the gap in the continuum part of the spectrum at every $k$ and the
existence of quasiparticle states outside the continuum spectrum and within
the gap. To show these features a very small broadening of $\eta =0.0001$
was used in panel (b).

\section{Bound states and band gaps}

\label{BoundSt}

The existence of a quasiparticle state in the Majumdar-Ghosh model (i.e.\ a
bound state below the continuum) was proved by ED in Ref.\ 
\cite{Maekawa} by
an analysis of the system size dependence of the pole strength $Z_h$
corresponding to the lowest eigenvalue. An analogous way was used to show
the existence of a quasiparticle in the single spinon approximation \cite
{Hayn} and in the diagrammatic approach \cite{Brenig01}. It was found that
the single spinon approximation gives $Z_h$ approximately twice as big as
that of the ED. That puzzle can be resolved when one takes into account the
linear growth of the recursion coefficients which does not show up in the
single spinon approximation. The linear growth may have an influence on the
weight $Z_h$. 
From a physical point of view the quasiparticle in this model represents the
bound state of holon and spinon. Scattering processes with the contribution
of high energy multispinon excited states that are not taken into account in
the single spinon approximation enlarge the probability of decay of the
bound state.

Fig.~7 shows the relative weights of the hole wave function
decomposition over recursion vectors $|u_n(k)\rangle$ for the three lowest
eigenenergies (thick solid, thin solid and dashed lines correspondingly) at
the band minimum $k=\pi /2$ . 90 recursive coefficients were taken: (a)
'bulk-related' ED coefficients, extrapolated by linear low; (b) single
spi\-non approximation. A qualitatively different behaviour is seen for the 
spin-polaron quasiparticle wave function (thick line) and the wave functions
that belong to the continuum. The different large-$n$ behaviour of recursion 
coefficients leads to different radius of the polaron state. 
For comparison, the
dotted line on panel (b) shows the change in the wave function caused by
the addition of the linear growth (\ref{asab}) to the single spinon 
approximation
coefficients. The linear growth leads to a larger radius of the polaron
state and to a decrease of the weight $Z_h$ by nearly a factor of two.
\begin{figure*}
\resizebox{0.45\textwidth}{!}{  \includegraphics{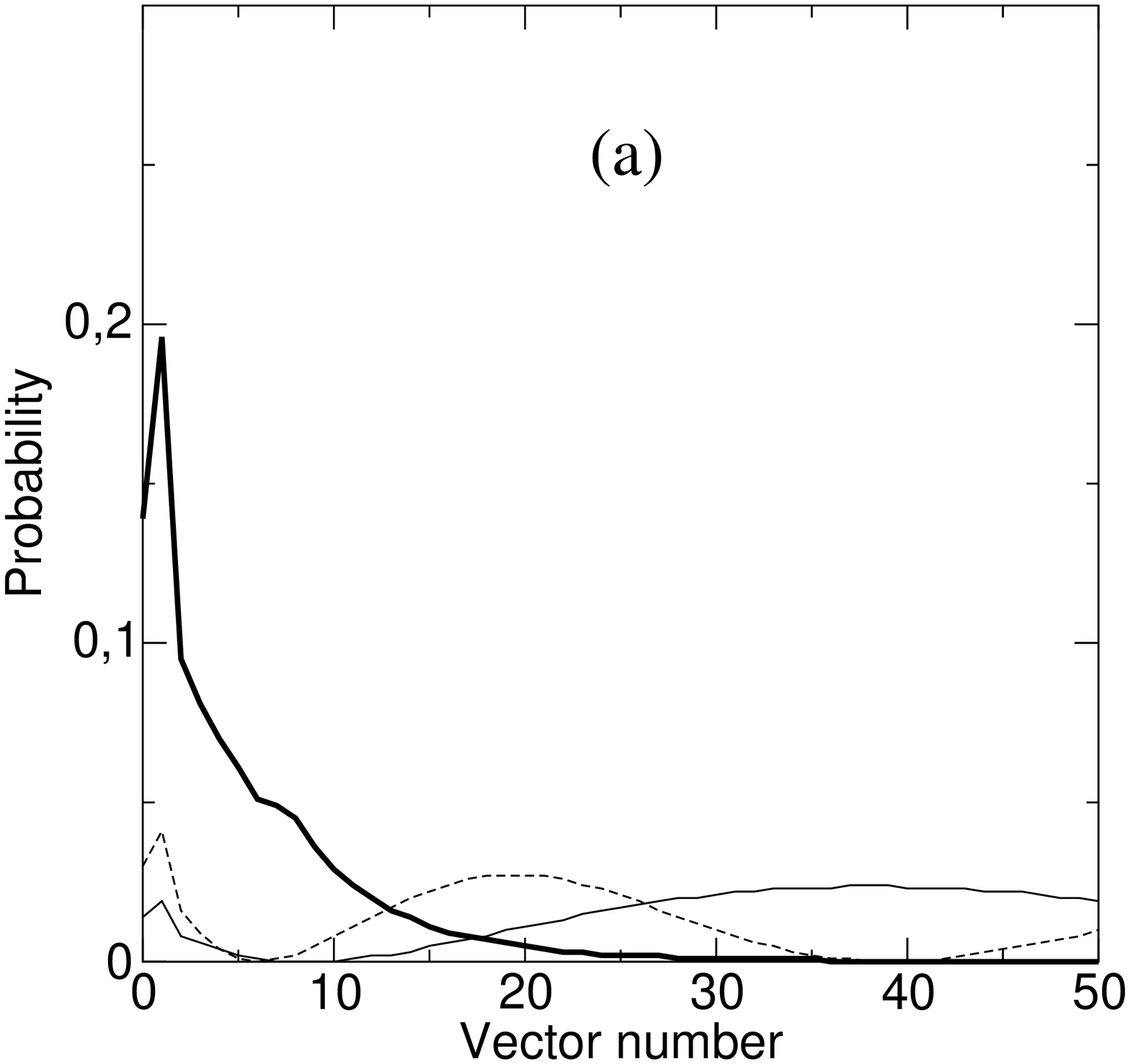} }
\resizebox{0.45\textwidth}{!}{  \includegraphics{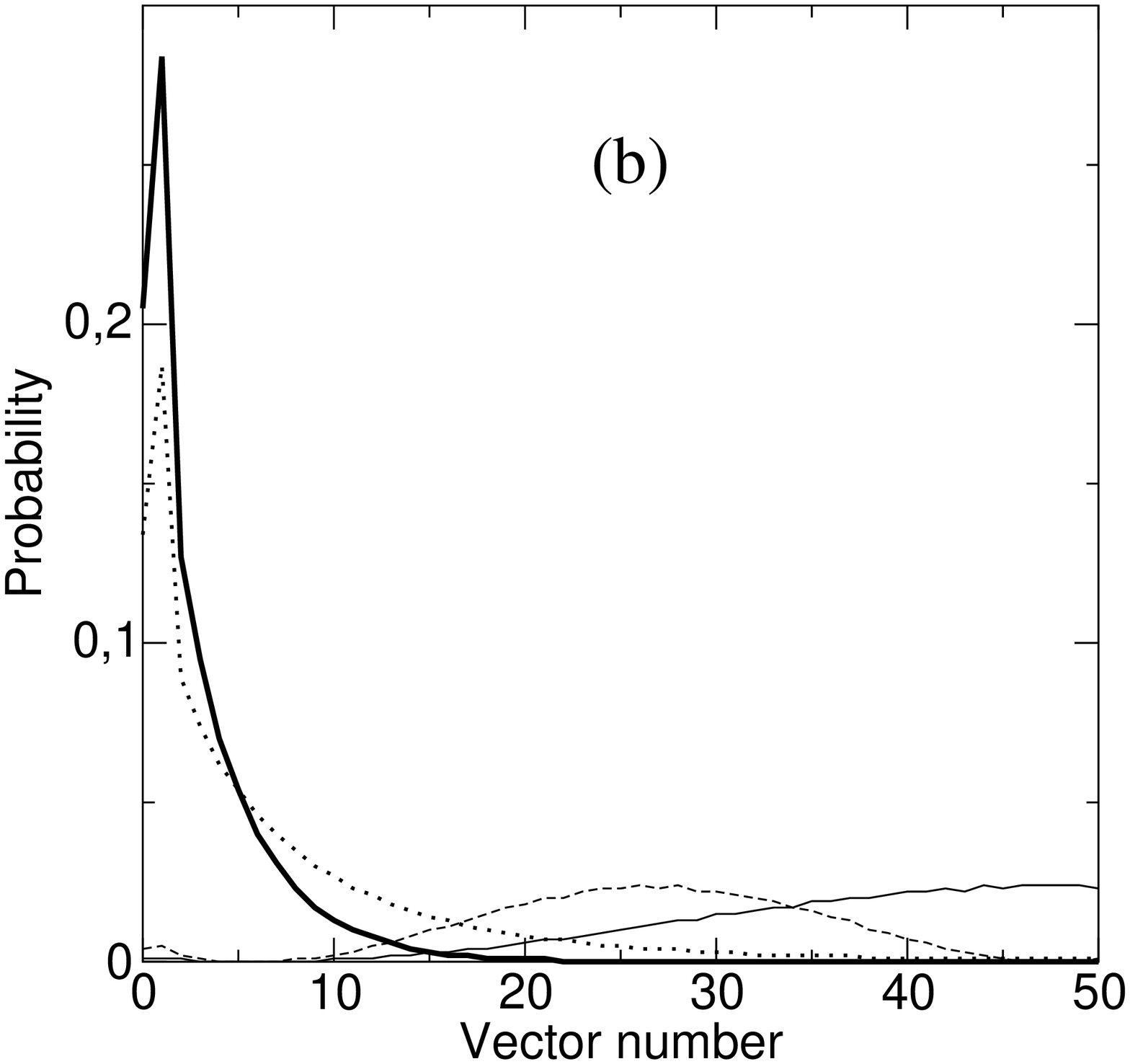} }
\caption{
The relative weights of the wave function decomposition over recursion 
vectors $|u_n(k)\rangle$ for the three lowest 
eigenenergies (thick solid, thin solid and
dashed lines, correspondingly) at the band minimum $k=\pi /2$.
90 recursive coefficients were taken: (a) 'bulk-related' ED coefficients, 
extrapolated by the linear law; (b) single spinon approximation. A 
qualitatively 
different behaviour is seen for the spin-polaron quasiparticle wave function 
(thick line) and the wave 
functions that belong to the continuum. The different large-$n$ behaviour of 
the recursion 
coefficients leads to different radii of the polaron state. For comparison, 
the dotted line on panel (b) shows the change in the wave function caused 
by the addition of the linear growth to the coefficients.
}
\end{figure*}

The physical origin of the gap in the continuum part of the spectrum that is
evident in the single spinon approximation may be understood as follows.
Let us recall that in the single spinon approximation we diagonalise the
Hamiltonian (\ref{H}) within the subspace spanned by the operator basis (\ref
{vkr}). (See the details in Sec.\ V.B and Appendix C of 
Ref.\ \cite{Hayn}.)
The Green's function (\ref{G}) for every $k$ is found from the solution of a
generalised eigenvalue problem that formally resembles an effective
tight-binding model (that is different for every $k$ and is determined by
the spin background) with Hamiltonian $E_{r,r^{\prime }}(k)$ in a 
non-orthogonal
basis with overlap matrix $S_{r,r^{\prime }}(k)$. The index $r$ 
denotes the distance between
'spinon' and 'holon' in (\ref{vkr}). The spectral density (\ref{A}) formally
coincides with the density of states of the effective model. In the limit 
$J,J^{\prime }\rightarrow 0$ both matrices $E$ and $S$ depend only on $
r-r^{\prime }$ and the effective model is 'translationally invariant' if we
consider $r$ and $r^{\prime }$ as site indices of the effective model. And we
obtain the results (\ref{Gt},\ref{ar}) making the Fourier 
transformation over 
$r-r^{\prime }$. This is the consequence of the degeneracy of all spin states
which makes it impossible for the holon to scatter on inhomogeneities of 
the spin state.
By switching on finite $J$ and $J^{\prime }$ the matrix $E_{r,r^{\prime
}}(k)$ acquires two contributions. One of them corresponds to the appearance of
the spinon coherent motion with the dispersion 
$\epsilon _s(Q)=-2J\cos Q$. This
contribution is also 'translationally invariant' and leads to changes in
the spectral density (\ref{ar}) described by (\ref{es}). Another contribution
comes from the commutation of the 'holon end' of the operator (\ref{vkr}) with
the spin part of the Hamiltonian (\ref{H}). It describes two processes, 
the loss of magnetic energy due to the presence of the holon, and 
the holon-spinon scattering. The former participates in the bound
state formation.  The latter has a contribution that is different for odd and
even $r-r^{\prime }$ , i.e.\ we observe a 'period doubling' in the effective
Hamiltonian. This is intimately connected with the dimerised nature of the 
Majumdar-Ghosh wave function. It is interesting that this feature does not
lead to a real period doubling in the system, i.e.\ the Green's function (\ref
{G}) and the spectral density (\ref{A}) have the whole Brillouin zone
periodicity, as we clearly see in Fig.~6.

\section{Summary}

We propose a new way how to extract the information about the infinite
system from the exact diagonalisation of small clusters. It is based on the
consideration of Lanczos recursion coefficients that are provided by ED. We
have found that the comparison of the results for several cluster sizes allows
to obtain the set of the coefficients that are not affected by the finite
size effects. These coefficients contain the needed information about the 
macroscopic system. When it is possible to infer their asymptotic behaviour
(that may strongly differ for the 'bulk related part' compared to the rest of
the set for a finite system) we propose to restore the Green's function and
spectral density using the terminator technique \cite{HayNex85}. This
may considerably improve the shape of the spectral density compared to the 
direct ED result which gives the spectral function in the form of a set of 
delta
functions.

\section*{Acknowledgements}

This work was supported by  DFG (436 UKR 113/49/41) and by a 
NATO Collaborative
Linkage Grant (PST.CLG. 976416). R.O.K. thanks for hospitality the IFW Dresden,
where the main part of this work has been carried out. For several numerical
calculations we used the Cambridge Recursion Library \cite{Nex84}. 


\end{document}